\documentclass{article}
\usepackage{spconf,amsmath,graphicx}

\usepackage{cite}

\usepackage{amssymb}
\usepackage{xcolor}
\usepackage{hhline}

\usepackage{hyperref}
\usepackage{url}

\usepackage{multirow} 
\usepackage{booktabs}

\title{EFFECTS OF LOMBARD REFLEX ON THE PERFORMANCE OF DEEP-LEARNING-BASED AUDIO-VISUAL SPEECH ENHANCEMENT SYSTEMS}
\name{Daniel Michelsanti$^1$, Zheng-Hua Tan$^1$, Sigurdur Sigurdsson$^2$, Jesper Jensen$^{1,2}$}
\address{$^1$ Aalborg University, Department of Electronic Systems, Denmark\\
	$^2$ Oticon A/S, Denmark\\
	\{danmi,zt,jje\}@es.aau.dk $\;$\{ssig,jesj\}@oticon.com}
\begin{document}
\ninept
\maketitle
\begin{abstract}
Humans tend to change their way of speaking when they are immersed in a noisy environment, a reflex known as Lombard effect. Current speech enhancement systems based on deep learning do not usually take into account this change in the speaking style, because they are trained with neutral (non-Lombard) speech utterances recorded under quiet conditions to which noise is artificially added. In this paper, we investigate the effects that the Lombard reflex has on the performance of audio-visual speech enhancement systems based on deep learning. The results show that a gap in the performance of as much as approximately 5 dB between the systems trained on neutral speech and the ones trained on Lombard speech exists. This indicates the benefit of taking into account the mismatch between neutral and Lombard speech in the design of audio-visual speech enhancement systems.
\end{abstract}
\section{Introduction}
\label{sec:intro}

Background noise can make vocal communication hard, because it degrades the speech of interest. However, speakers instinctively react to the presence of background noise and change their speaking style to maintain their speech intelligible. This reflex is known as \textit{Lombard effect} \cite{brumm2011evolution}, and it is characterised, acoustically, by an increase in speech sound level \cite{junqua1993lombard}, a longer word duration \cite{pittman2001recognition}, and modifications of the speech spectrum \cite{junqua1993lombard}, and, visually, by a speech hyper-articulation \cite{garnier2018hyper}. 

In particularly challenging situations, e.g.~when the listener is hearing impaired and the noise level is high, this natural change of speaking style might not suffice to guarantee an effective communication. Hence, there is a need to reduce the negative effects of background noise on speech quality and intelligibility with speech enhancement (SE) techniques. SE is important in several applications, and proposed approaches range from classical statistical model-based methods \cite{loizou2013speech}, to deep-learning-based ones \cite{wang2018supervised}. These techniques use only audio signals to perform enhancement, and we refer to them as audio-only SE (AO-SE) systems.

During speech production, the movements of some articulatory organs, e.g.~lips and jaw, might be visible to the listener, enhancing speech perception \cite{erber1975auditory, sumby1954visual}. Exploiting the information conveyed by these visual cues, which are independent of the acoustical environment where SE systems operate, leads to systems that are more robust than the AO-SE ones to background noise. This has already been shown in early work on audio-visual SE (AV-SE) \cite{girin2001audio}. More complex frameworks have been proposed later, e.g.~\cite{almajai2011visually}, in which a voice activity detector and phoneme-specific methods are used to estimate noise and clean speech statistics for a visually derived Wiener filter. Very recently, deep-learning-based techniques have also been adopted to solve the AV-SE task \cite{hou2017audio, gabbay2017visual, ephrat2018looking, afouras2018conversation}.

AV-SE systems are likely to be deployed in acoustic situations where AO-SE systems underperform or fail, e.g. in situations where the background noise level is high and the Lombard effect is clearly present. In other words, the typical input to AV-SE systems is Lombard speech in noise. However, existing SE systems usually ignore this effect, being trained with clean speech signals recorded in quiet to which noise is artificially added. The mismatch between the neutral and the Lombard speaking styles can lead to sub-optimal performance of audio-only-based speaker \cite{hansen2009analysis} and speech recognition \cite{junqua1993lombard} systems. Only a few works investigate the impact of the Lombard effect on visual \cite{heracleous2013analysis, marxer2018impact} and audio-visual \cite{heracleous2013analysis} automatic speech recognition, but, to the best knowledge of the authors, no studies have been conducted for AV-SE systems. 

The aim of the current paper is to examine to which extent a deep-learning-based AV-SE system trained on neutral speech can effectively enhance Lombard speech. This is important to understand, because if such a system can model well Lombard speech, then there is no need to include in the training procedure speech recorded in Lombard conditions, which is usually hard to obtain. Specifically, we are interested in answering the following research questions:
\begin{enumerate}
	\item Is an AV-SE system trained on neutral speech able to improve Lombard speech?
	\item Does a performance gap exist between a system trained on Lombard speech and a system trained on neutral speech when tested on Lombard speech from speakers that have been observed during training (seen speakers)?
	\item Is a performance gap still present for speakers that have not been observed during training (unseen speakers)?
\end{enumerate}
The last two questions are relevant to understand the impact of inter-speaker differences of Lombard speech on SE. We expect that the system trained on Lombard speech  enhances the Lombard speech of a seen speaker better than the system trained on neutral speech, because it should model well the Lombard speaking style of that speaker. However,  the system trained on Lombard speech may have difficulties in generalising to unseen speakers, because the characteristics of the Lombard speech of one person might significantly differ from the characteristics of the Lombard speech of another person \cite{junqua1993lombard, marxer2018impact}.



%
%
%

\section{Audio-visual corpus and noise data}
\label{sec:audio-visual corpus and noise data}

The dataset used in this study is the English language Lombard GRID corpus \cite{alghamdi2018corpus}, consisting of audio and visual (frontal and profile view\footnote{In this study, the audio and the frontal view video recordings are used.}) recordings from 54 subjects (24 males and 30 females). The audio and video channels are temporally aligned. Each speaker is recorded, while pronouncing 50 unique six-word sentences, whose syntax is identical to the one in GRID \cite{cooke2006audio}, in each of two conditions: non-Lombard (NL) and Lombard (L). In the condition NL, the speakers are recorded with a microphone placed at 30 cm in front of their mouth and two cameras mounted on a helmet worn by them. The condition L replicates the same setup, but it simulates the Lombard effect by presenting speech shaped noise (SSN) at a level of 80 dB sound pressure level (SPL) through headphones. In addition, the speakers are provided with a carefully adjusted self-monitoring feedback, while reading aloud some sentences to a listener, who asks to repeat the utterances from time to time in order to simulate possible miscomprehensions of speech in noise. This scenario allows to take into account the two factors responsible for the Lombard adaptation: first, speakers tend to regulate their vocal effort based on the auditory feedback, i.e.~they involuntarily react to the perceived level of their own speech \cite{marxer2018impact}; secondly, they change their speaking style to communicate better with others \cite{lane1971lombard, lu2008speech}.

%
%
%
%
%
%

The impact of the noise type on the Lombard effect is currently unclear. While some studies have found no evidence to support a systematic active response of speakers to the spectral characteristics of the noise \cite{lu2009speech, garnier2014speaking}, Hansen and Varadarajanare \cite{hansen2009analysis} indicate the presence of differences across noise types in the way that Lombard effect occurs. Following this finding, we use SSN, since this is the noise type that was presented to the speakers of the Lombard GRID corpus. The noise was generated as reported in \cite{kolboek2016speech}.

The audio-visual corpus and the noise data are used to build training, validation and test sets as explained in Sec. \ref{sec:experiments}.

\section{Methods}
\label{sec:methods}

The goal of many SE systems is to estimate the clean signal $x(n)$, given a mixture $y(n)=x(n)+d(n)$, where $d(n)$ is an additive noise signal and $n$ denotes the discrete-time index. Usually, the SE problem is tackled in the time-frequency (TF) domain, where the additive noise model is expressed as $Y(k,l)=X(k,l)+D(k,l)$, with $k$ indicating the frequency bin index, $l$ denoting the time frame index, and $Y(k,l)$, $X(k,l)$, and $D(k,l)$ being the short-time Fourier transform (STFT) coefficients of $y(n)$, $x(n)$, and $d(n)$, respectively. Since the estimation of the phase of the clean STFT coefficients, $X(k,l)$, with a neural network is hard \cite{williamson2016complex}, enhancement can be performed by estimating $A_{k,l}=\mathopen| X(k,l) \mathclose|$ from $R_{k,l}=\mathopen| Y(k,l) \mathclose|$. The time-domain signal is obtained using the estimated clean magnitude spectrum and the noisy phase in an inverse STFT procedure.

In this study, we use a mask approximation (MA) approach, where a neural network is trained to learn the ideal amplitude mask (IAM), defined as $M^{\text{IAM}}_{k,l}~=~\frac{A_{k,l}}{R_{k,l}}$, with the following objective function:
\begin{align}J = \frac{1}{TF} \sum_{k,l} \big( M^{\text{IAM}}_{k,l} - \widehat{M}_{k,l}\big)^2, \label{eqn:IAM-MA}\end{align}
where $\widehat{M}_{k,l}$ is the output of the network, $k \in \{1, \ldots , F\}$, $l \in \{ 1, \ldots , T\}$, and $TF$ is the size of the training target. This objective function showed best performance in several conditions in a comparison study \cite{michelsanti2019training}, where a range of targets and cost functions, used to train a deep-learning-based AV-SE system, are investigated.

\subsection{Preprocessing}
\label{subsec:audio preprocessing}

The audio signals, which have a sample rate of 16 kHz, are peak-normalised to 1 per signal. Then, a 640-point STFT is applied, using a 640-sample-long Hamming window and a hop size of 160 samples. Due to spectral symmetry, we consider only the 321 bins that cover the positive frequencies.

To preprocess the video signals, which have a frame rate of 25 fps, we make use of the detection and alignment algorithms implemented in the dlib toolkit \cite{dlib09}. In particular, for each frame we detect the face with a linear support vector machine (SVM) on histogram of oriented gradients (HOG) features, and track it across the frames with a Kalman filter. Then, the detected face is aligned using 5 landmark points that identify the corners of the eyes and the bottom of the nose and scaled to 256~$\times$~256 pixels. Finally, the 128~$\times$~128-pixel region around the mouth is extracted.

\subsection{Architecture and training procedure}
\label{subsec:architecture}

The neural network architecture, inspired by \cite{gabbay2017visual} and identical to \cite{michelsanti2019training}, consists of four blocks: a video encoder, an audio encoder, a fusion subnetwork, and an audio decoder.

The video encoder takes as input 5 consecutive grayscale frames of the mouth region, corresponding to 200 ms. Six convolutional layers are applied, and each of them is followed by: leaky-ReLU activation, batch normalisation, 2$\times$2 strided max-pooling with a 2$\times$2 kernel, and dropout with a probability of 25\%. The audio encoder is fed with a 20-frame-long magnitude spectrogram, corresponding to 200 ms, and consists of 6 convolutional layers, followed by leaky-ReLU activation and batch normalisation. Further details regarding the convolutional layers of the encoders are shown in Table~\ref{tab:convenc}. The inputs of the encoders are normalised to have zero mean and unit variance based on the training set statistics.

\begin{table}[ht]
\centering
\caption{Convolutional layers of the audio and video encoders. For the video encoder, a 1$\times$1 stride is always applied.}
\label{tab:convenc}
\begin{tabular}{c | c c | c c c }
\toprule
                           \multicolumn{1}{c}{}   & \multicolumn{2}{c}{Video Encoder} & \multicolumn{3}{c}{Audio Encoder} \\
\midrule
{Layer} & \# Filters   & Kernel   & \# Filters   & Kernel   & Stride   \\
\midrule
{1}      &      128        &    5$\times$5      &     64          &   5$\times$5       &      2$\times$2      \\ 
{2}      &      128        &    5$\times$5      &     64          &   4$\times$4       &      2$\times$1     \\ 
{3}      &      256        &    3$\times$3      &     128        &   4$\times$4       &      2$\times$2    \\ 
{4}      &      256        &    3$\times$3      &     128        &   2$\times$2       &      2$\times$1   \\ 
{5}      &      512        &    3$\times$3      &     128        &   2$\times$2       &      2$\times$1    \\ 
{6}      &      512        &    3$\times$3      &     128        &   2$\times$2       &      2$\times$1     \\ 
\bottomrule
\end{tabular}
\end{table}

The outputs of the encoders are concatenated and fed into the fusion subnetwork, consisting of 3 fully connected layers, the first 2 with 1312 neurons and the last one with 3840. A leaky-ReLU activation is used for all the layers.

The audio decoder takes as input the result vector of the fusion subnetwork and processes it with 6 transposed convolutional layers that mirror the audio encoder ones. The architecture has 3 skip connections between layers 1, 3, and 5 of the audio encoder and the corresponding layers of the decoder. The output layer uses ReLU activation. In the end, a 321$\times$20 mask matrix, which estimates the IAM, is obtained. The target values are clipped between 0 and 10 \cite{wang2014training}. 

After the initialisation of the weights with the Xavier approach, the network is trained for 50 epochs adopting the Adam optimiser, with the objective function in Eq.~(\ref{eqn:IAM-MA}), a batch size of 64 and an initial learning rate of $4\cdot10^{-4}$. The network is evaluated on the validation set every epoch, and the learning rate is halved, if the validation loss increases. For testing, we use the network that performs the best on the validation set across the 50 epochs to avoid overfitting issues.

Besides this AV-SE system, we also train an AO-SE and a video-only SE (VO-SE) architectures, obtained by removing the video encoder or the audio encoder, respectively, from the AV-SE system.

\subsection{Audio-visual speech enhancement}
\label{subsec:postprocessing}

The enhancement of the noisy signals is performed in three steps. First, the preprocessed non-overlapping audio and video (or only one of the two modalities when AO-SE and VO-SE architectures are used) sequences are forward propagated through the network. Then, the resulting masks $\widehat{M}_{k,l}$ are concatenated and point-wise multiplied with the complex-valued STFT spectrogram of the mixture. Finally, the enhanced signals are reconstructed with the inverse STFT.

\section{Experiments}
\label{sec:experiments}

This section describes the experimental setup and the evaluation measures used in this study. The training, validation and test data have been allocated differently between the seen and the unseen speaker cases, since the amount of speech material available to train deep-learning-based systems is relatively small.

\subsection{Seen speaker case}

Each person may exhibit a different Lombard speaking style \cite{junqua1993lombard, marxer2018impact}. Modelling these differences could be performed by training several speaker-dependent SE systems. However, this choice requires a large amount of speech data for each speaker. Instead, we adopt a multi-speaker setup and train one AV-SE system with 54 speakers for each of the two conditions of the database, L and NL, obtaining two models, AV-L and AV-NL, respectively. 

For AV-L, the utterances of the database recorded in condition L are randomly shuffled and organised into: a test set with 10 utterances for each speaker; a validation set consisting of 5 utterances per speaker; and a training set with the remaining material. 

For AV-NL, the training and validation sets are arranged by picking the neutral utterances corresponding to the Lombard utterances used for the training and validation of AV-L. The test set is the same as the one used for AV-L, because we are interested in investigating the enhancement potential of AV-NL in condition L and compare it with the AV-L performance. We also train two VO-SE systems, for the conditions L and NL, and two AO-SE systems, for the conditions L and NL, obtaining four additional models: VO-L, VO-NL, AO-L, and AO-NL, respectively. This should be considered as an additional aspect to the research questions introduced in Sec.~\ref{sec:intro}, which allows us to understand the contribution of each modality to the enhancement of Lombard speech. 

\subsection{Unseen speaker case}

Generalisation of SE systems to unseen speakers is important, especially in applications where it is hard to collect speech data to train a speaker-dependent system. For this reason, we want to examine whether a system trained with Lombard speech can generalise well, i.e.~better than a system trained on neutral speech, to the Lombard speech of an unseen speaker. We perform a 6-fold cross-validation by training 6 AV-SE systems on utterances in condition L (5 per speaker for validation and the rest for training) of 45 speakers, and testing them on utterances in condition L of 9 unseen speakers (4 males and 5 females). We refer to each of these models as AV-L*. The same procedure is applied for training and validation of 6 AV-SE systems using the corresponding utterances recorded in condition NL. The obtained models are denoted as AV-NL*.

All the models used in this study are summarised in Table \ref{tab:models}.

\begin{table}[ht]
\centering
\caption{Models used in this study for the seen and the unseen (indicated with a *) speaker cases.}
\label{tab:models}
\begin{tabular}{c | c | c }
\toprule

&\multicolumn{2}{c}{Training Material}\\

Modality & Non-Lombard Speech & Lombard Speech \\
\midrule

Vision     		&    VO-NL & VO-L        \\ 
Audio      		&    AO-NL & AO-L       \\ 
Audio-visual     	&    AV-NL / AV-NL* & AV-L / AV-L*    \\ 

\bottomrule

\end{tabular}
\end{table}

\subsection{Additive noise levels}

To construct the training, the validation, and the test sets for all the models, the speech signals from the Lombard GRID database are mixed with additive SSN at 6 signal to noise ratios (SNRs), in uniform steps between -20 dB and 5 dB. The SNR range has been chosen due to the following considerations:
\begin{enumerate}
	\item Current SE systems are trained on noisy signals in which noise is added to the clean signals at several SNRs to ensure robustness to different noise levels. For this reason, we do not train SNR-specific systems.
	\item In the Lombard GRID database, the energy difference between Lombard and neutral utterances is between 3 dB and 13 dB \cite{marxer2018impact}. If we assume that the listener is immersed in SSN at 80 dB SPL, like in the recording conditions of the Lombard GRID database, and that the conversational speech level is between 60 and 70 dB SPL \cite{raphael2007speech, moore2012introduction}, the SNR is between -17 dB and 3 dB. The slightly wider SNR range used in the experiments (between -20 dB and 5 dB) is chosen to take into account the possible speech level variations due to the distance of the listener from the speaker.
\end{enumerate}

\subsection{Evaluation metrics}

The performance of all the models are evaluated in terms of two objective measures, namely perceptual evaluation of speech quality (PESQ) \cite{rix2001perceptual} as implemented in \cite{loizou2013speech}, and extended short-time objective intelligibility (ESTOI) \cite{jensen2016algorithm}, because they are good estimators of speech quality and intelligibility, respectively. PESQ ranges from -0.5 to 4.5, where high values correspond to high speech quality. For ESTOI, whose range is practically between 0 and 1, higher scores correspond to higher speech intelligibility.

\section{Results and Discussion}
\label{sec:results}

\begin{figure}[!h]
  \centering
  \includegraphics[width=0.49\textwidth]{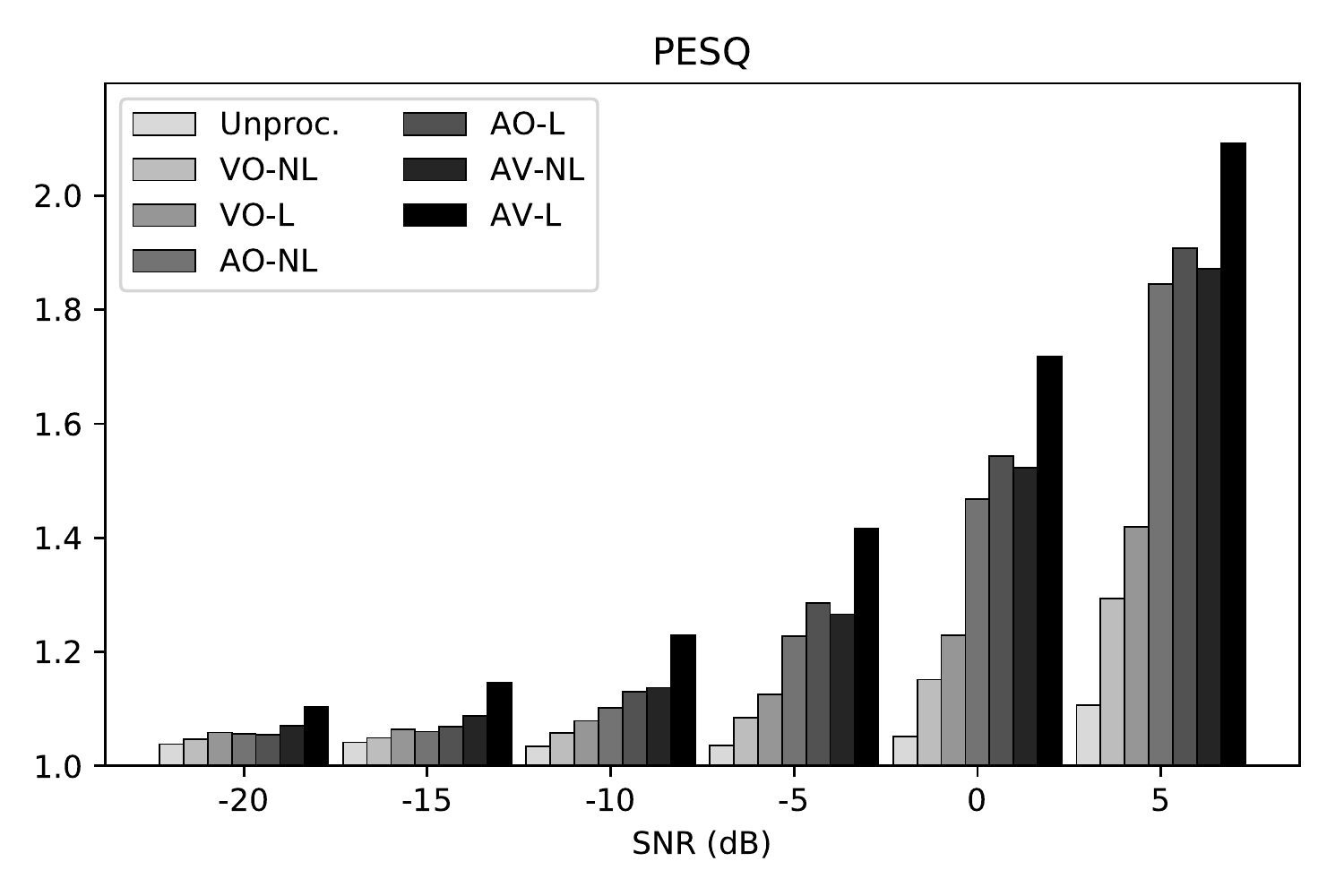}
  \caption{PESQ scores for the seen speaker case. \textit{Unproc.} refers to the unprocessed signals.}
  \label{fig:PESQ_seen}
\end{figure}

\begin{figure}[!h]
  \centering
  \includegraphics[width=0.49\textwidth]{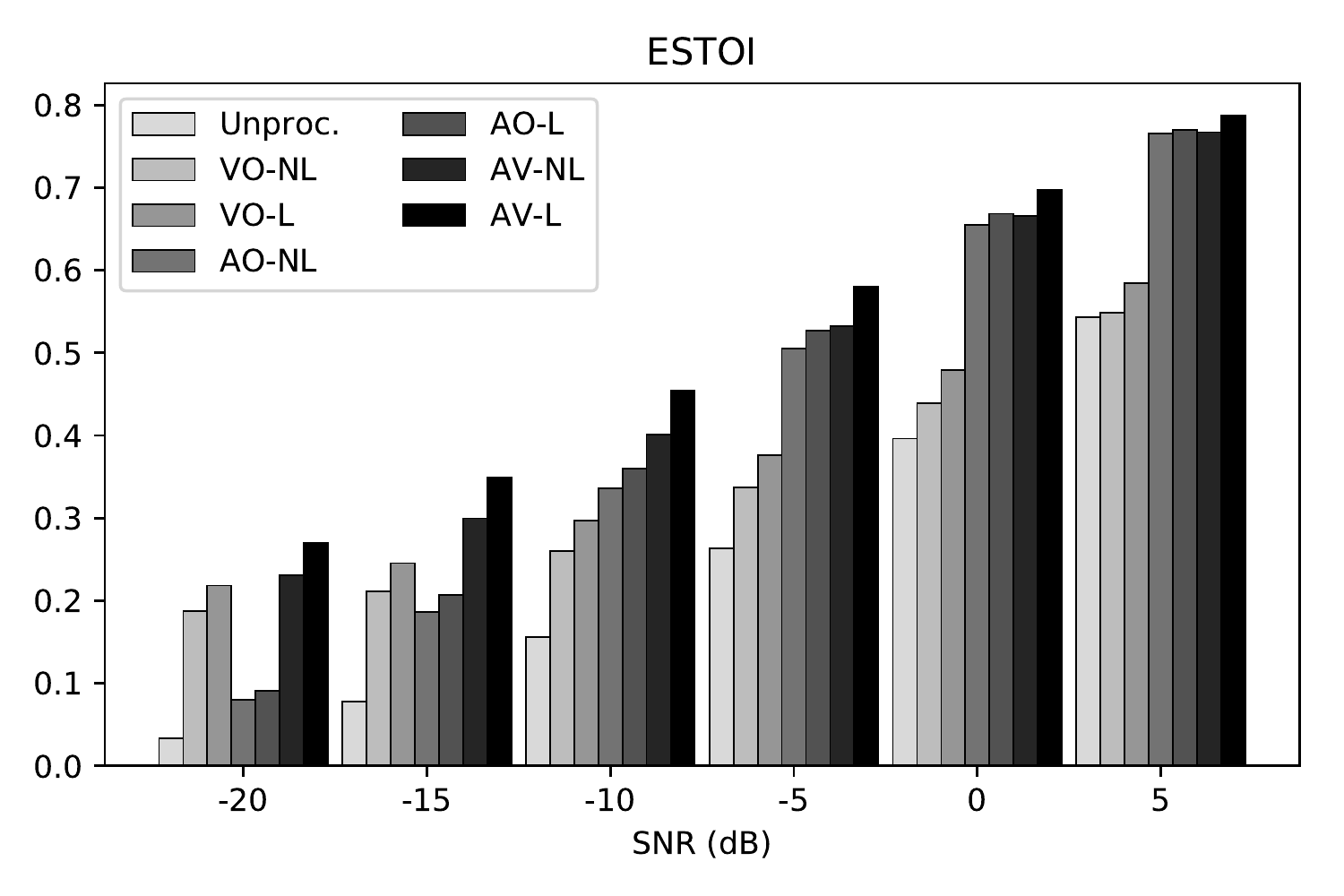}
  \caption{ESTOI scores for the seen speaker case. \textit{Unproc.} refers to the unprocessed signals.}
  \label{fig:ESTOI_seen}
\end{figure}

Figs.~\ref{fig:PESQ_seen} and \ref{fig:ESTOI_seen} show the PESQ and the ESTOI scores, respectively, for the seen speaker case. It can be seen that all the models improve the mixtures in terms of both estimated speech quality and intelligibility at all SNRs.

Regarding PESQ (Fig.~\ref{fig:PESQ_seen}), AV-L performs slightly better than AV-NL at -20 dB SNR, but the gap between the two models become larger when the SNR increases. The PESQ performance of the AV-L system at a particular SNR is almost as high as that of the AV-NL system at an SNR of 5 dB higher. This makes it clear that training a model with speech recorded in condition L is beneficial. The VO-L and the AO-L systems also outperform their NL counterparts, with a smaller performance gap.

The ESTOI performance (Fig.~\ref{fig:ESTOI_seen}) shows a similar trend, with L models outperforming the corresponding NL ones. In this case, the performance difference is substantial even at very low SNRs, where the SNR gain as defined above is slightly less than 5 dB. As expected, the contribution of vision to intelligibility is higher at low SNRs. Interestingly, the gap between VO-NL and VO-L is larger than the one between AO-NL and AO-L, suggesting that visual differences between the two speaking styles have a higher impact on intelligibility enhancement than acoustic differences.

\begin{figure}[!h]
  \centering
  \includegraphics[width=0.49\textwidth]{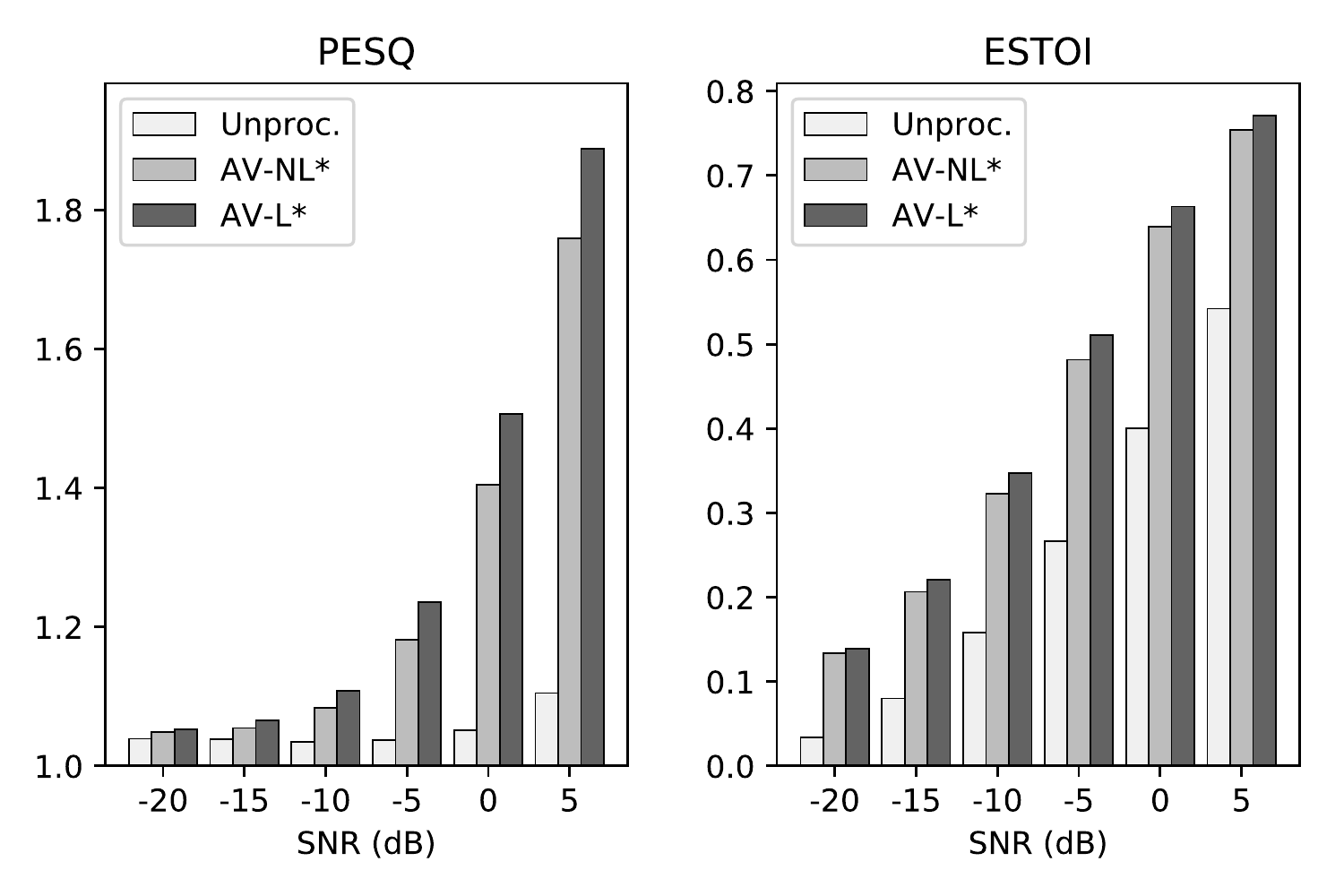}
  \caption{PESQ and ESTOI scores for the unseen speaker case. \textit{Unproc.} refers to the unprocessed signals.}
  \label{fig:unseen}
\end{figure}

The results for the unseen speaker case are shown in Fig.~\ref{fig:unseen}. The performance of AV-L* is always better than AV-NL* at all the SNRs in terms of both PESQ and ESTOI. As observed in the seen speaker case, the major PESQ improvements are reported at high SNRs, where the performance gap between AV-NL* and AV-L* is substantial. Regarding ESTOI, the difference between AV-NL* and AV-L* is smaller than the one observed in the seen speaker case. This can be explained by potential difficulties in modelling the inter-speaker variations of the Lombard speaking style. However, the performance gap between the two models is still evident, especially between -10 dB and 0 dB SNRs, indicating the benefit of using Lombard speech for training.

\section{Conclusion}
\label{sec:conclusion}

This paper investigated the impact of Lombard effect on audio-visual speech enhancement. For this purpose, the Lombard GRID database containing the recordings of 54 different speakers in both Lombard and non-Lombard conditions has been used to train and test deep-learning-based speech enhancement systems.
From the results of the experiments, the following conclusions can be drawn:
\begin{enumerate}
	\item A network trained on neutral speech is able to improve noisy Lombard speech in terms of both estimated speech quality and intelligibility.
	\item When the models are evaluated on seen speakers, the gap in the performance between the systems trained on neutral speech and the ones trained on Lombard speech indicates a benefit of as much as 5 dB if Lombard speech is used during training.
	\item When the models are evaluated on unseen speakers, the performance difference between the systems trained on neutral speech and the systems trained on Lombard speech is smaller than the one observed in the seen speakers scenario, but it still suggests the advantage of training speech enhancement systems with Lombard speech.
\end{enumerate}
This study showed that the Lombard effect has an impact on the performance of audio-visual speech enhancement systems and that the mismatch between neutral and Lombard speech should be taken into account in the design of these systems. Future works include listening tests to confirm the findings obtained with objective measures of speech quality and speech intelligibility.

\bibliographystyle{IEEEbib}
\bibliography{icassp2}

\begin{thebibliography}{10}

\bibitem{brumm2011evolution}
H.~Brumm and S.~A. Zollinger,
\newblock ``The evolution of the {Lombard} effect: 100 years of psychoacoustic
  research,''
\newblock {\em Behaviour}, vol. 148, no. 11-13, pp. 1173--1198, 2011.

\bibitem{junqua1993lombard}
J.-C. Junqua,
\newblock ``The {Lombard} reflex and its role on human listeners and automatic
  speech recognizers,''
\newblock {\em The Journal of the Acoustical Society of America}, vol. 93, no.
  1, pp. 510--524, 1993.

\bibitem{pittman2001recognition}
A.~L. Pittman and T.~L. Wiley,
\newblock ``Recognition of speech produced in noise,''
\newblock {\em Journal of Speech, Language, and Hearing Research}, vol. 44, no.
  3, pp. 487--496, 2001.

\bibitem{garnier2018hyper}
M.~Garnier, L.~M{\'e}nard, and B.~Alexandre,
\newblock ``Hyper-articulation in {Lombard} speech: An active communicative
  strategy to enhance visible speech cues?,''
\newblock {\em The Journal of the Acoustical Society of America}, vol. 144, no.
  2, pp. 1059--1074, 2018.

\bibitem{loizou2013speech}
P.~C. Loizou,
\newblock {\em Speech enhancement: theory and practice},
\newblock CRC press, 2013.

\bibitem{wang2018supervised}
D.~L. Wang and J.~Chen,
\newblock ``Supervised speech separation based on deep learning: An overview,''
\newblock {\em IEEE/ACM Transactions on Audio, Speech, and Language
  Processing}, 2018.

\bibitem{erber1975auditory}
N.~P. Erber,
\newblock ``Auditory-visual perception of speech,''
\newblock {\em {Journal of Speech and Hearing Disorders}}, vol. 40, no. 4,
  1975.

\bibitem{sumby1954visual}
W.~H. Sumby and I.~Pollack,
\newblock ``Visual contribution to speech intelligibility in noise,''
\newblock {\em The Journal of the Acoustical Society of America}, vol. 26, no.
  2, 1954.

\bibitem{girin2001audio}
L.~Girin, J.-L. Schwartz, and G.~Feng,
\newblock ``Audio-visual enhancement of speech in noise,''
\newblock {\em The Journal of the Acoustical Society of America}, vol. 109, no.
  6, pp. 3007--3020, 2001.

\bibitem{almajai2011visually}
I.~Almajai and B.~Milner,
\newblock ``Visually derived {Wiener} filters for speech enhancement,''
\newblock {\em IEEE Transactions on Audio, Speech, and Language Processing},
  vol. 19, no. 6, pp. 1642--1651, 2011.

\bibitem{hou2017audio}
J.-C. Hou, S.-S. Wang, Y.-H. Lai, Y.~Tsao, H.-W. Chang, and H.-M. Wang,
\newblock ``Audio-visual speech enhancement using multimodal deep convolutional
  neural networks,''
\newblock {\em IEEE Transactions on Emerging Topics in Computational
  Intelligence}, vol. 2, no. 2, pp. 117--128, 2018.

\bibitem{gabbay2017visual}
A.~Gabbay, A.~Shamir, and S.~Peleg,
\newblock ``Visual speech enhancement,''
\newblock in {\em Proc. of Interspeech}, 2018.

\bibitem{ephrat2018looking}
A.~Ephrat, I.~Mosseri, O.~Lang, T.~Dekel, K.~Wilson, A.~Hassidim, W.~T.
  Freeman, and M.~Rubinstein,
\newblock ``Looking to listen at the cocktail party: A speaker-independent
  audio-visual model for speech separation,''
\newblock {\em ACM Transactions on Graphics}, vol. 37, no. 4, pp.
  112:1--112:11, 2018.

\bibitem{afouras2018conversation}
T.~Afouras, J.~S. Chung, and A.~Zisserman,
\newblock ``The conversation: Deep audio-visual speech enhancement,''
\newblock {\em Proc. of Interspeech}, 2018.

\bibitem{hansen2009analysis}
J.~H.~L. Hansen and V.~Varadarajan,
\newblock ``Analysis and compensation of {Lombard} speech across noise type and
  levels with application to in-set/out-of-set speaker recognition,''
\newblock {\em IEEE Transactions on Audio, Speech, and Language Processing},
  vol. 17, no. 2, pp. 366--378, 2009.

\bibitem{heracleous2013analysis}
P.~Heracleous, C.~T. Ishi, M.~Sato, H.~Ishiguro, and N.~Hagita,
\newblock ``Analysis of the visual {Lombard} effect and automatic recognition
  experiments,''
\newblock {\em Computer Speech \& Language}, vol. 27, no. 1, pp. 288--300,
  2013.

\bibitem{marxer2018impact}
R.~Marxer, J.~Barker, N.~Alghamdi, and S.~Maddock,
\newblock ``The impact of the {Lombard} effect on audio and visual speech
  recognition systems,''
\newblock {\em Speech Communication}, vol. 100, pp. 58--68, 2018.

\bibitem{alghamdi2018corpus}
N.~Alghamdi, S.~Maddock, R.~Marxer, J.~Barker, and G.~J. Brown,
\newblock ``A corpus of audio-visual {Lombard} speech with frontal and profile
  views,''
\newblock {\em The Journal of the Acoustical Society of America}, vol. 143, no.
  6, pp. EL523--EL529, 2018.

\bibitem{cooke2006audio}
M.~Cooke, J.~Barker, S.~Cunningham, and X.~Shao,
\newblock ``An audio-visual corpus for speech perception and automatic speech
  recognition,''
\newblock {\em The Journal of the Acoustical Society of America}, vol. 120, no.
  5, pp. 2421--2424, 2006.

\bibitem{lane1971lombard}
H.~Lane and B.~Tranel,
\newblock ``The {Lombard} sign and the role of hearing in speech,''
\newblock {\em Journal of Speech, Language, and Hearing Research}, vol. 14, no.
  4, pp. 677--709, 1971.

\bibitem{lu2008speech}
Y.~Lu and M.~Cooke,
\newblock ``Speech production modifications produced by competing talkers,
  babble, and stationary noise,''
\newblock {\em The Journal of the Acoustical Society of America}, vol. 124, no.
  5, pp. 3261--3275, 2008.

\bibitem{lu2009speech}
Y.~Lu and M.~Cooke,
\newblock ``Speech production modifications produced in the presence of
  low-pass and high-pass filtered noise,''
\newblock {\em The Journal of the Acoustical Society of America}, vol. 126, no.
  3, pp. 1495--1499, 2009.

\bibitem{garnier2014speaking}
M.~Garnier and N.~Henrich,
\newblock ``Speaking in noise: How does the {Lombard} effect improve acoustic
  contrasts between speech and ambient noise?,''
\newblock {\em Computer Speech \& Language}, vol. 28, no. 2, pp. 580--597,
  2014.

\bibitem{kolboek2016speech}
M.~Kolb{\ae}k, Z.-H. Tan, and J.~Jensen,
\newblock ``Speech enhancement using long short-term memory based recurrent
  neural networks for noise robust speaker verification,''
\newblock in {\em Proc. of SLT}, 2016.

\bibitem{williamson2016complex}
D.~S. Williamson, Y.~Wang, and D.~L. Wang,
\newblock ``Complex ratio masking for monaural speech separation,''
\newblock {\em IEEE/ACM Transactions on Audio, Speech and Language Processing},
  vol. 24, no. 3, pp. 483--492, 2016.

\bibitem{michelsanti2019training}
D.~Michelsanti, Z.-H. Tan, S.~Sigurdsson, and J.~Jensen,
\newblock ``{On training targets and objective functions for
  deep-learning-based audio-visual speech enhancement},''
\newblock {\em Submitted to ICASSP}, 2019.

\bibitem{dlib09}
D.~E. King,
\newblock ``Dlib-ml: A machine learning toolkit,''
\newblock {\em Journal of Machine Learning Research}, vol. 10, pp. 1755--1758,
  2009.

\bibitem{wang2014training}
Y.~Wang, A.~Narayanan, and D.~L. Wang,
\newblock ``On training targets for supervised speech separation,''
\newblock {\em IEEE/ACM Transactions on Audio, Speech and Language Processing
  (TASLP)}, vol. 22, no. 12, pp. 1849--1858, 2014.

\bibitem{raphael2007speech}
L.~J. Raphael, G.~J. Borden, and K.~S. Harris,
\newblock {\em Speech science primer: Physiology, acoustics, and perception of
  speech},
\newblock Lippincott Williams \& Wilkins, 2007.

\bibitem{moore2012introduction}
B.~C.~J. Moore,
\newblock {\em An introduction to the psychology of hearing},
\newblock Brill, 2012.

\bibitem{rix2001perceptual}
A.~W. Rix, J.~G. Beerends, M.~P. Hollier, and A.~P. Hekstra,
\newblock ``Perceptual evaluation of speech quality ({PESQ}) - a new method for
  speech quality assessment of telephone networks and codecs,''
\newblock in {\em Proc. of ICASSP}, 2001.

\bibitem{jensen2016algorithm}
J.~Jensen and C.~H. Taal,
\newblock ``An algorithm for predicting the intelligibility of speech masked by
  modulated noise maskers,''
\newblock {\em IEEE/ACM Transactions on Audio, Speech, and Language
  Processing}, vol. 24, no. 11, pp. 2009--2022, 2016.

\end{thebibliography}

\end{document}